\documentclass[%
12pt,
 superscriptaddress,
 amsmath,amssymb,
 aps,UTF8,
]{revtex4-1}

\usepackage{graphicx}
\usepackage{dcolumn}
\usepackage{bm}
\usepackage{xcolor}
\usepackage{url}
\usepackage{float}
\usepackage{lipsum}
\usepackage{array}

\usepackage{booktabs}

\usepackage{subfigure}
\usepackage{multirow}

\begin{document}


\title{ Proton Shell Structure }
\author{Wei Kou}
\email{kouwei@impcas.ac.cn}
\affiliation{Institute of Modern Physics, Chinese Academy of Sciences, Lanzhou 730000, China}
\affiliation{University of Chinese Academy of Sciences, Beijing 100049, China}

\author{Chengdong Han}
\email{chdhan@impcas.ac.cn}
\affiliation{Institute of Modern Physics, Chinese Academy of Sciences, Lanzhou 730000, China}
\affiliation{University of Chinese Academy of Sciences, Beijing 100049, China}

\author{Xurong Chen}
\email{Corresponding author: xchen@impcas.ac.cn}
\affiliation{Institute of Modern Physics, Chinese Academy of Sciences, Lanzhou 730000, China}
\affiliation{University of Chinese Academy of Sciences, Beijing 100049, China}
\affiliation{Guangdong Provincial Key Laboratory of Nuclear Science, Institute of Quantum Matter, South China Normal University, Guangzhou 510006, China}



\begin{abstract}
We study the internal structure of the proton and propose a shell structure model of the proton. Based on the mass distribution and mass radius of proton we analyze the experiments at different energy scales and conclude that there should be shell structure inside proton, and the quark and gluon shells are energy-scale-dependent. We successfully explain the dip-like structure in the proton-proton elastic scattering cross section distribution from TOTEM Collaboration using the mass decomposition of proton and the proton shell structure at different energy scales.

\end{abstract}

\maketitle


\section{Introduction}
\label{sec:intro}

The atom, which until more than 110 years ago was thought to be a substance with an evenly distributed charge, called ``plum pudding model" \cite{Thomson:1904bjw}. Towards the end of the first decade of the 20th century, Ernest Rutherford and his two students performed the famous experiment of alpha particles hitting gold foil target \cite{Rutherford:1911zz}. The results of experiment were at odds with common sense, and surprisingly there was alpha particles scattering with large angle. It is suggested us that the distribution of positive and negative charges within the atom is uneven, and that the positive charges should be concentrated at the center of atom. This new model provides novel ideas for future investigations into the structure of atom. Scientists believed that matter is divisible, atom have nucleus inside them, and it was later discovered that nucleus also have structure, consisting of proton and the neutral nucleon -- neutron \cite{Chadwick:1932ma}.

Proton, as a nucleon, is one of the basic units of the nucleus. Properties such as mass and charge of proton have been studied for almost 100 years. Due to confinement effects in quantum chromodynamics (QCD), understandings of the structure inside proton have remained very poor for several years. In order to experimentally determine the dynamic information inside proton, Deep Inelastic Scattering (DIS) experiments have been designed. By hitting  proton target with high-energy electron or photon, physicists found which was involved at the reaction was actually unit so-called ``parton" inside proton. The parton model describes the data from high-energy scattering experiments very well, it is one of the many methods used to probe the internal structure of proton. Although it is theoretically possible to research the internal components of proton through experimental data and model building, the few data and large errors of experiments at the low-energy scales make it impossible to get high-precision tests and predictions. Due to the upgrading of the experimental setup in recent years, with the maturity of high energy colliders, there is an urgent need for lower energy colliders to ensure that the proton is not broken up during the scattering process. After thorough researches physicists believe that the Electron-Ion Collider \cite{Accardi:2012qut,Chen:2018wyz,Chen:2020ijn,AbdulKhalek:2021gbh,Anderle:2021wcy} can achieve this goal and can better approximate real physics with high statistical precision data.

The origin of the mass of the proton, one of the most specific issues in nature of the proton, has been the hot spot people discussed. The mass of the valence quark given by the Higgs mechanism \cite{Higgs:1964pj} is only about one percent of the whole proton mass, and the source of mass is, according to modern physics, closely related to the gluons inside proton. For this reason, many physics researchers have fully explored the issues related to the mass of proton \cite{Ji:1995sv,Guo:2021ibg,Hatta:2018ina,Hatta:2019lxo,Wang:2019mza,Ji:2021mtz,Ji:2021pys,Ji:2021qgo,Kharzeev:2021qkd,Kou:2021bez,Kou:2021qdc,Liu:2021gco,Lorce:2017xzd,Metz:2020vxd,Rodini:2020pis,Chen:2020gml,Wang:2021dis}. The gravitational effect inside proton would be translated into work on solving the trace of energy-momentum tensor under the weak gravitational field approximation \cite{Kharzeev:2021qkd}. Based on this method it is possible to give information on the gravitational form factor from experimental data on near-threshold vector meson photo-production processes and thus get the mass radius of the proton \cite{Kharzeev:2021qkd,Wang:2021dis} as well as mass distribution. The proton mass distribution reveals the structure information of proton. There is no doubt that the internal mass of the proton is not homogeneous, so the exploration of the structure inside proton will help us to deepen our understandings of the proton structure.

Since most of the source of mass within proton comes from the behavior of gluon, it is particularly important to study the dynamic properties of gluon. In the case of electron (photon)-proton scattering at high energy, the gluons are dominant at the Bjorken scale $x_B<10^{-2}$ case. Vector meson production and Deep Virtual Compton Scattering (DVCS) \cite{Ji:1996nm} processes can occur in this energy region and can be explained by the dipole model. Theoretically, it is possible to transfer the cross section calculation from the QCD factorization \cite{Collins:1996fb} to the dipole cross section, where the expression for the dipole cross section contains information about gluons dynamical properties of proton. In Ref. \cite{Kowalski:2003hm,Kowalski:2006hc,Caldwell:2010zza,Mantysaari:2016ykx,Mantysaari:2016jaz}, authors analyzed the experimental data on high-energy $\gamma p$ scattering using dipole representation and gave the sizes of transverse gluons in proton according to the IP-Sat model \cite{Kowalski:2003hm}. The behavior of the valence quark within proton can be described by the axial form factor \cite{Frankfurt:2002ka}. The neutrino nucleon scattering experiment is generally used to extract the nucleon axial form factor and axial charge \cite{MiniBooNE:2010bsu,Meyer:2016oeg}. Of course, Lattice QCD (LQCD) \cite{Alexandrou:2017hac,Jang:2019vkm,RQCD:2019jai} and Dyson-Schwinger equation (DSE) methods \cite{Chen:2020wuq} can be used to do the same research theoretically.

The background of the current study and experimental measurements or analysis of parton were discussed in previous paragraphs. Due to the running nature of the strong interaction coupling constants, the parton distribution functions in proton are renormalization-energy-scale-dependent. Then the behavior of the corresponding gluons as well as valence quarks should also be scale-dependent. On the contrary, as fundamental properties of proton, charge and mass must be independent of energy scales, which are fundamental physical quantities. In order to give a self-consistent and reasonable picture of the multi-shell structure inside proton, we take the charge and mass radius of the proton and their corresponding distributions as criteria, and finally gave a distributions of the parts inside proton and their corresponding radii, starting from the available experimental data as much as possible.

The purpose of this article is to provide a novel image of proton internal structure -- shell structure.  Sec. \ref{sec:Radii} gives information on the different components or distributions of proton for each radius and a brief analysis.  Sec. \ref{sec:multilyer} gives an image of the multi-shell structure inside proton based on the mass radius as well as the mass distribution of the proton from the experiment. The mass distribution and mechanical properties of proton will be talked in Sec. \ref{sec:mechanical}. In Sec. \ref{sec:exp} we briefly talk about the recent experiment by TOTEM Collaboration in 2019 and simple phenomenological explanation of proton shell structure. Finally we make some brief discussions and conclusions.

\section{Proton internal distributions and Radii}
\label{sec:Radii}
There are four fundamental interactions in nature as we know them, electromagnetic, weak, strong and gravitational interactions. These four fundamental interactions can be used to detect corresponding properties in proton. Take the measurement of the electromagnetic properties of proton as an example, physicists can use the charged leptons at a certain energy to hit the proton target, and determine the electromagnetic interaction between the charged leptons and proton by observing the momentum transfer of the emitted charged leptons. This idea should be applied to other interactions as well. The form factor is a very convenient tool to probe the internal structure of proton, which can be measured by experiments to get more accurate results.
We state that the result of the radius calculation for the proton depends on the dipole form factor, which can be written as
\begin{equation}
	G(t) = \frac{G(0)}{\left(1-\frac{t}{\Lambda^2}\right)^2},
	\label{eq:dipole form}
\end{equation}
where $\Lambda$ can be determined by experimental data. The proton root-mean-square radius is usually determined by the following form
\begin{equation}
	G(t)=1+\frac{1}{6}\left\langle r_{p}^{2}\right\rangle t+O\left(t^{2}\right),
	\label{eq:radii calculations}
\end{equation}
it can be got the root-mean-square radius by neglecting the second order term as follows,
\begin{equation}
	\left\langle r_{p}^{2}\right\rangle=\left.\frac{6}{G(0)} \cdot \frac{\mathrm{d}G(t)}{\mathrm{d} t}\right|_{t=0}.
	\label{eq:differ}
\end{equation}
If the components of the proton are not unique, then for different components corresponds to different distributions. We assume here that the proton is spherically symmetric. The distribution of gluons could be given from DIS or DVCS experiments. The structure of gluons during diffraction process at high energy is dependent on scale. Therefore the fitted slope $B$ of differential cross-section by experiments according to the exponential form $\mathrm{d}\sigma/\mathrm{d}t = A\exp(-Bt)$ should also be energy-dependent \cite{Kowalski:2006hc,Caldwell:2010zza}. In these works we mentioned earlier, the authors compared the experimental data of HERA for various periods and used the dipole model to give the three-dimensional gluon radius in $\gamma p$ scattering at a center-of-mass energy of $\sqrt{s} = 30$ GeV, and the radius is assigned a value $r_{2g} = 0.61\pm0.04$ fm \cite{Caldwell:2010zza}. 

Valence quarks are the source of gluons released by high-energy proton. As discussed in the previous section, the valence quark is responsible for the charged part of the proton, producing charge distribution from the center and gradually weakening along the radial direction. In addition to the charge nature, quarks are also involved in strong interactions and, under certain conditions, in weak interactions. The valence quark distribution inside proton is determined by the axial form factor \cite{Frankfurt:2002ka}, and currently the best way to measure the axial form factor as well as the axial charge is through neutrino-nucleon scattering experiments. In Ref. \cite{MiniBooNE:2010bsu} the MiniBooNE Collaboration gives the cutting parameter of axial form factor by fitting experimental data read as $M_A = 1.35\pm 0.17$ GeV, corresponding to an axis radius of $r_A = 0.506\pm 0.064$ fm from dipole assumption \cite{Frankfurt:2002ka}.

\begin{figure}[H]
	\centering
	\includegraphics[width=0.6\textwidth]{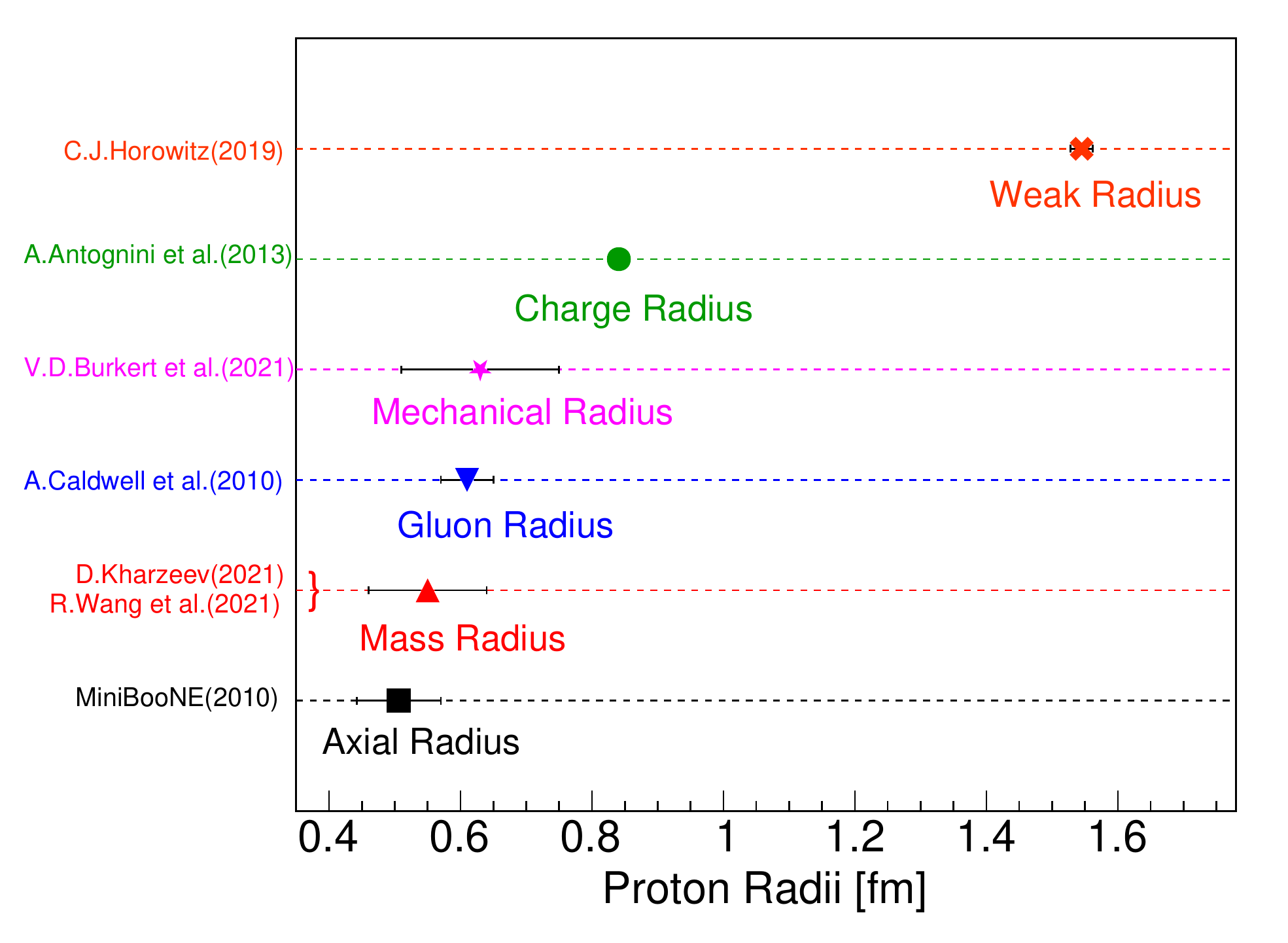}
	\caption{Comparison between different radii from the experimental data and analysis \cite{MiniBooNE:2010bsu,GlueX:2019mkq,Kharzeev:2021qkd,Wang:2021dis,Caldwell:2010zza,Burkert:2021ith,Antognini:2013txn,Horowitz:2018yxh}. Charge and weak radii measurements are not shown because the uncertainty is too small.
	}
	\label{fig:radii-compare}
\end{figure}

The mass and charge of proton are the fundamental physical quantities. In recent years, there have been numerous results on the mass of proton. In the previous work \cite{,Kharzeev:2021qkd}, author used the data from the J/$\psi$ photo-production experiment at the near threshold \cite{GlueX:2019mkq} to combine with the energy-momentum tensor of QCD and finally give the dipole parameter $M_s =1.24\pm0.07$ GeV of the gravitational form factor. The author gave the corresponding mass radius $r_m = 0.55\pm 0.03$ fm of proton, too. Meanwhile, we also proposed the similar results \cite{Wang:2021dis}. Finally we simply quote the charge radius results for the proton obtained from the muonic hydrogen Lamb shift experiment, $r_c = 0.8409\pm0.0004$ fm \cite{Antognini:2013txn,ParticleDataGroup:2020ssz}. Not only proton has electromagnetic and gravitational properties, it can also couple with $Z^0$ bosons, and the quantity that characterizes the strength of their coupling is called the weak charge of the proton. The $Q_{weak}$ Collaboration at Jefferson Lab has also made high-precision measurements of the weak charge of proton \cite{Qweak:2013zxf,Qweak:2018tjf}. In a recent research work \cite{Horowitz:2018yxh}, the author proposed a method for calculating the weak radius of proton and gave results for $r_w = 1.545\pm0.017$ fm is 80$\%$ larger than charge radius. The mechanical properties inside proton are reflected in the D-term of the gravitational form factor. The pressure and shear force distribution within the proton can be described by the D form factor. The DVCS experiment \cite{Burkert:2021ith} can obtain the information of D-term and the mechanical radius of the proton \cite{Polyakov:2018zvc}. The size of mechanical radius is about 75$\%$ of the charge one.  To compare the various radii of proton we discussed, we organize the above results and discussions into FIG. \ref{fig:radii-compare}. For ease of reading we also give the results in tabular form, see TABLE. \ref{tab:radii} for details.

\begin{table}[H]
	\centering
	\caption{Various radii of proton.}
	\begin{center}
		\begin{tabular}{ |c|c|c|c|c|c|c| }
			\toprule[1.3pt]
			Radius & Axial & Mass  & Gluon&Mechanical& Charge& Weak\\
			\hline
			$r$ (fm) & $0.506\pm0.064$ & $0.55\pm0.09$ & $0.61\pm0.04$&$0.63\pm0.12$ &$0.8409\pm0.0004$&$1.545\pm0.017$\\
			
			\bottomrule[1.3pt]
		\end{tabular}	
		\label{tab:radii}
	\end{center}
\end{table}

\section{Multi-shell distribution of proton}
\label{sec:multilyer}
\begin{widetext}
\begin{figure}[htbp]
	\centering
	\subfigure[]{
		\label{fig:massdist}
		\includegraphics[width=0.42\textwidth]{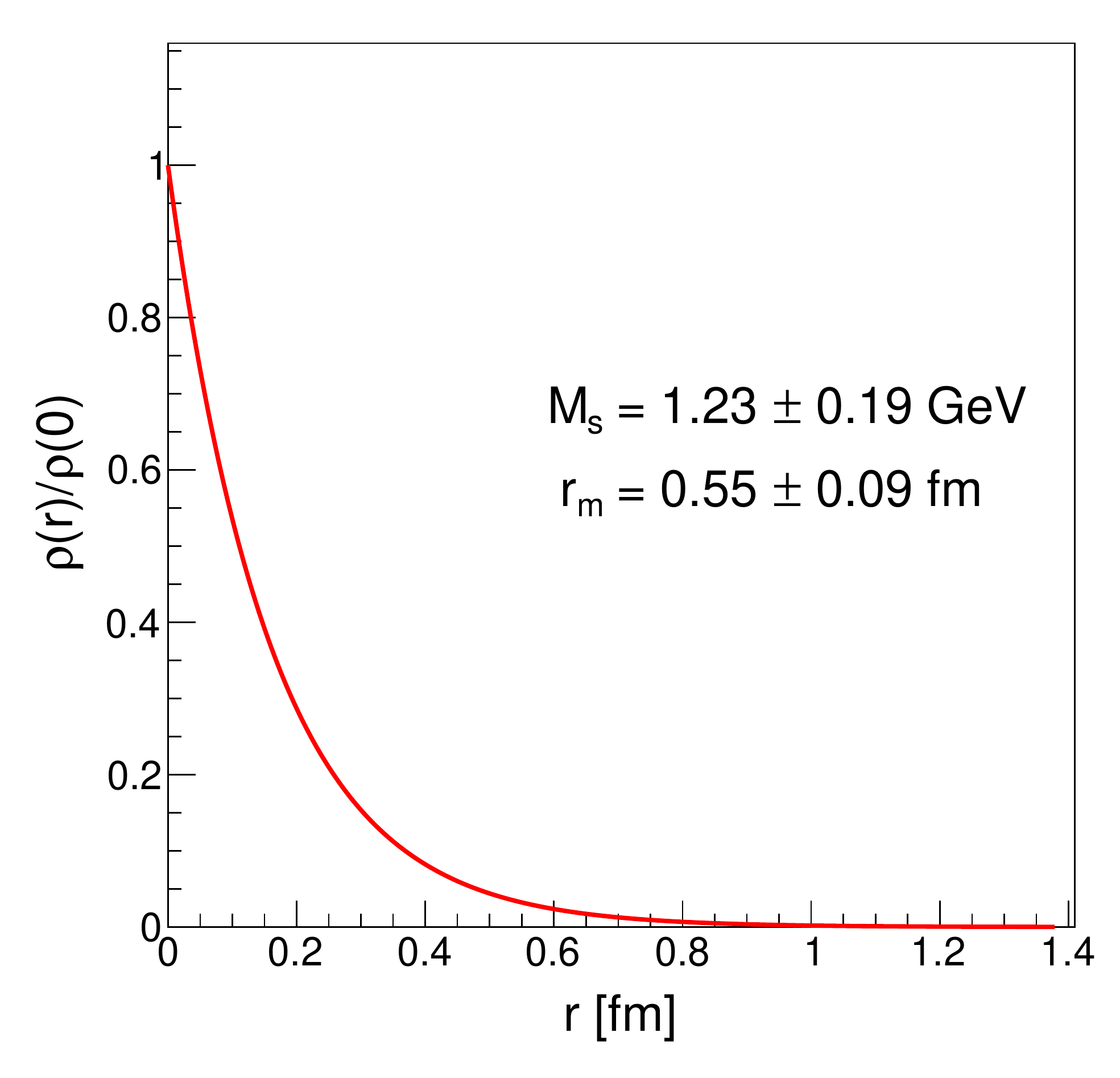}}
	\subfigure[]{
		\label{fig:layer}
		\includegraphics[width=0.45\textwidth]{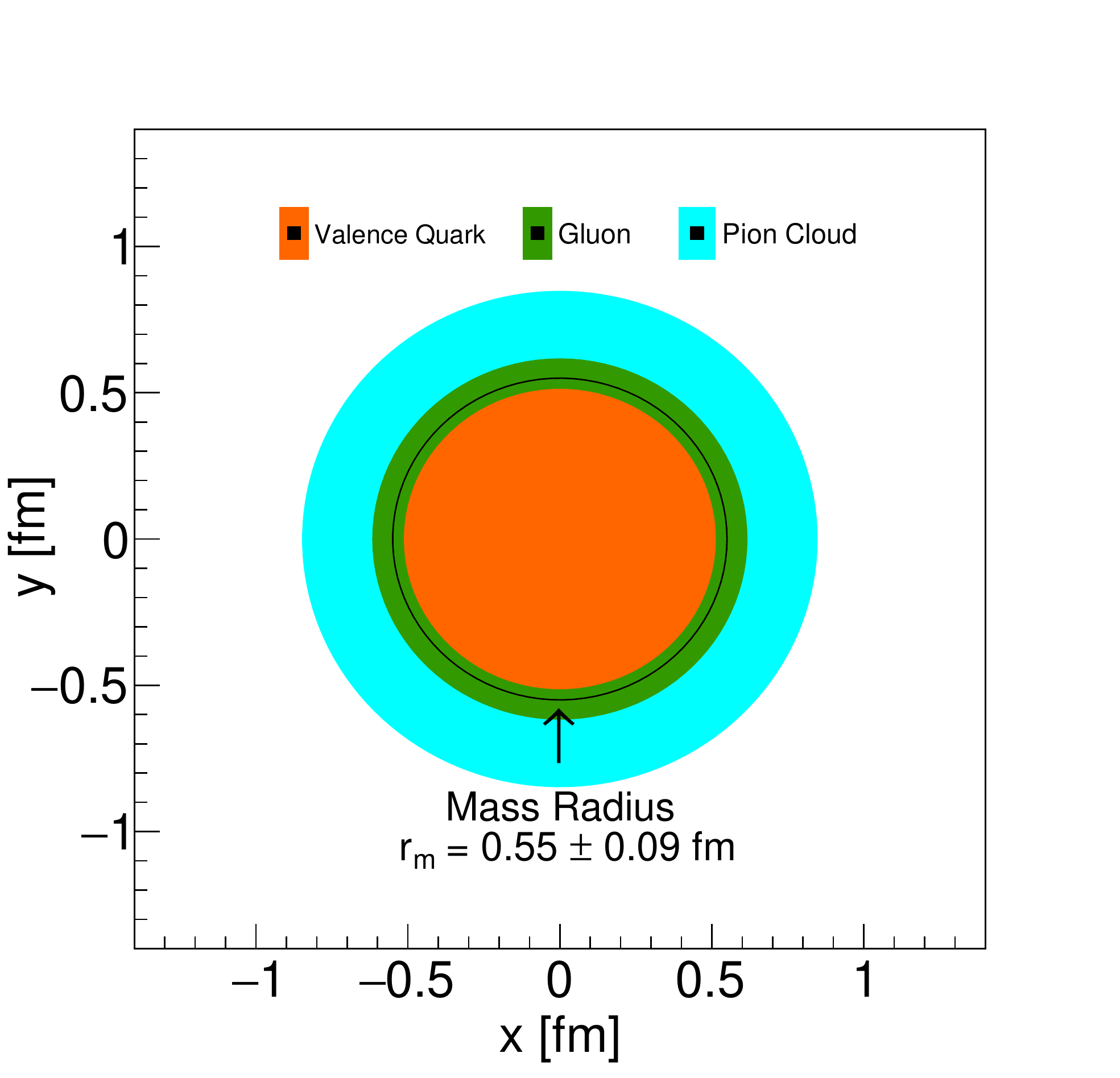}}
	\caption{(Color online) (a) The mass density distribution described by gravitational form factor. (b) The schematic diagram of the proton multi-shell structure at low-$Q^2$. The orange region depicts the distribution of valence quarks inside proton, corresponding to the axial radius, and the green region is the region of gluons produced from the valence quarks. Valence quarks and gluons account for the largest mass fraction of proton. Black solid line represents the mass radius. In the outermost layer is the structure of the pion cloud. It is important to emphasize that we take the charge radius as the boundary limit of proton.
	}
	\label{fig:schematic}
\end{figure}
\end{widetext}

In order to study the internal layered structure of the proton, we review the earlier studies related to the proton radius. Inspired by Ref. \cite{Dremin:2018orv,Dremin:2018urc,Dremin:2019pza,Dremin:2019swd,Islam:2005by,Islam:2008sfk}, we discuss in this section some new methods for dividing proton. Based on the sizes of the various radii of the proton discussed earlier, we sort them along the proton radial direction, i.e. $r_A<r_m<r_{2g}<r_c$. With this relationship, we can construct the multi-shell structure inside proton. We use the charge radius as the maximum boundary value for the radius of the proton. First, the proton has smallest value of the axis radius, so the region closest to the centre is the valence quark distribution of the proton. The distribution varies with the energy scale, but not by much. The maximum density of radial valence quarks is located near the axial radius of the proton, $r_A=0.506$ fm. Second, at the low energy ($Q^2$) and small Bjorken scale $x_B<10^{-2}$ condition, the gluon generated by the valence quark which can be regarded as sources and roughly described as diffusing out in the form of Gaussian wave packet. The half-width of the Gaussian-type wave packet can be thought of the gluon radius $r_{2g}=0.61$ fm of proton which evolves with the energy scale. Charge distribution is unaffected during the discussion. According to Ref. \cite{Strikman:2003gz}, the formation of quark-antiquark pairs from gluons splitting in vacuum can be regarded as pion cloud structure at the proton surface. We have represented the above text in simple diagrams (See FIG. \ref{fig:schematic}). FIG. \ref{fig:massdist} comes from the previous works such as \cite{GlueX:2019mkq,Kharzeev:2021qkd,Wang:2021dis}. FIG. \ref{fig:layer} reveals the multi-shell structure of the proton based on the mass radius and mass distribution of the proton and different experimental data, which appears to be quite clear. In Sec. \ref{sec:mechanical} we will discuss the physical image of this diagram corresponding the mechanical information and explain its plausibility.

\newpage

\section{Mechanical properties inside proton}
\label{sec:mechanical}
\begin{figure}[htbp]
	\centering  
	\subfigure[]{
		\label{fig:1d}
		\includegraphics[width=0.42\textwidth]{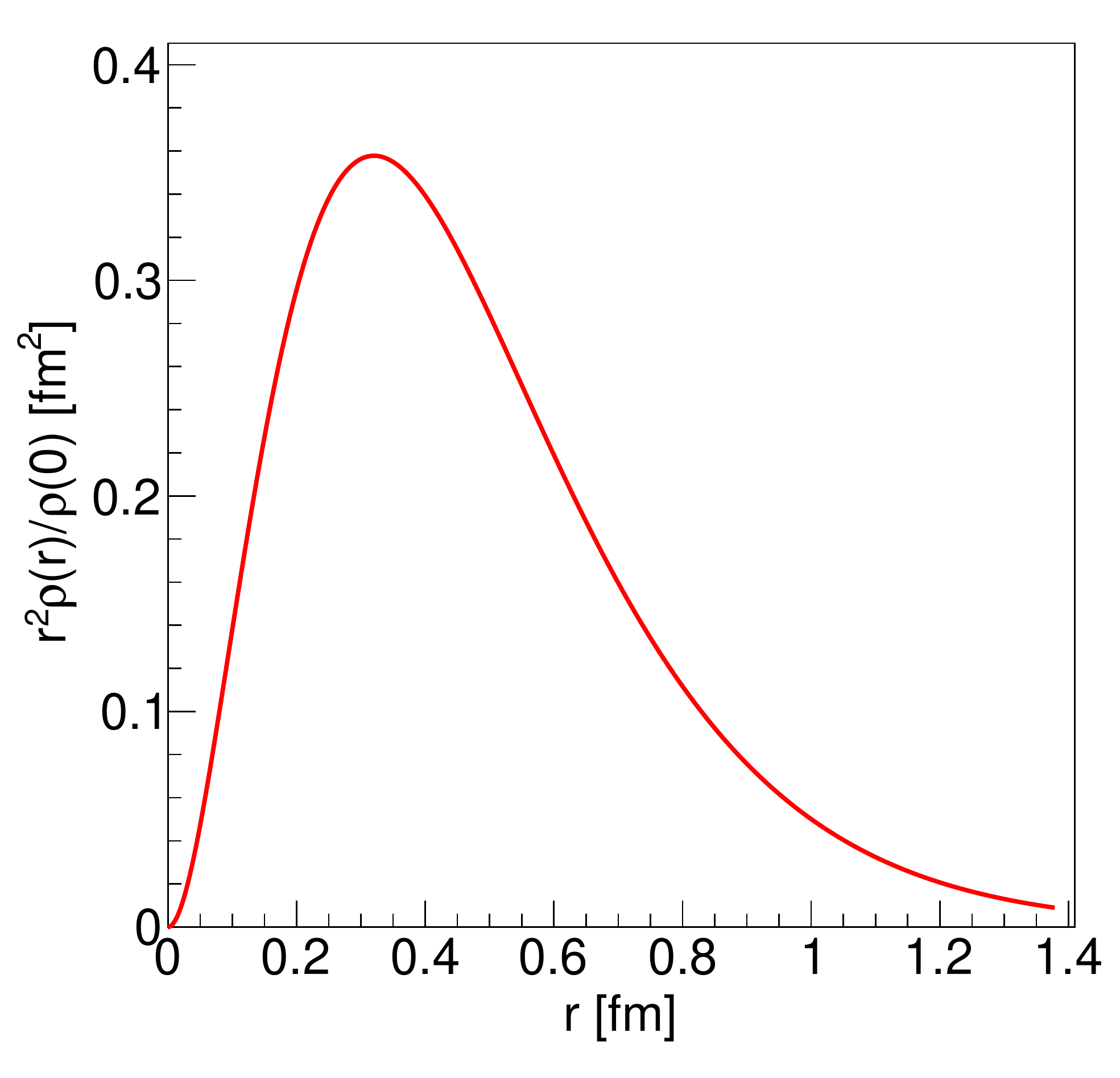}}
	\subfigure[]{
		\label{fig:2d}
		\includegraphics[width=0.45\textwidth]{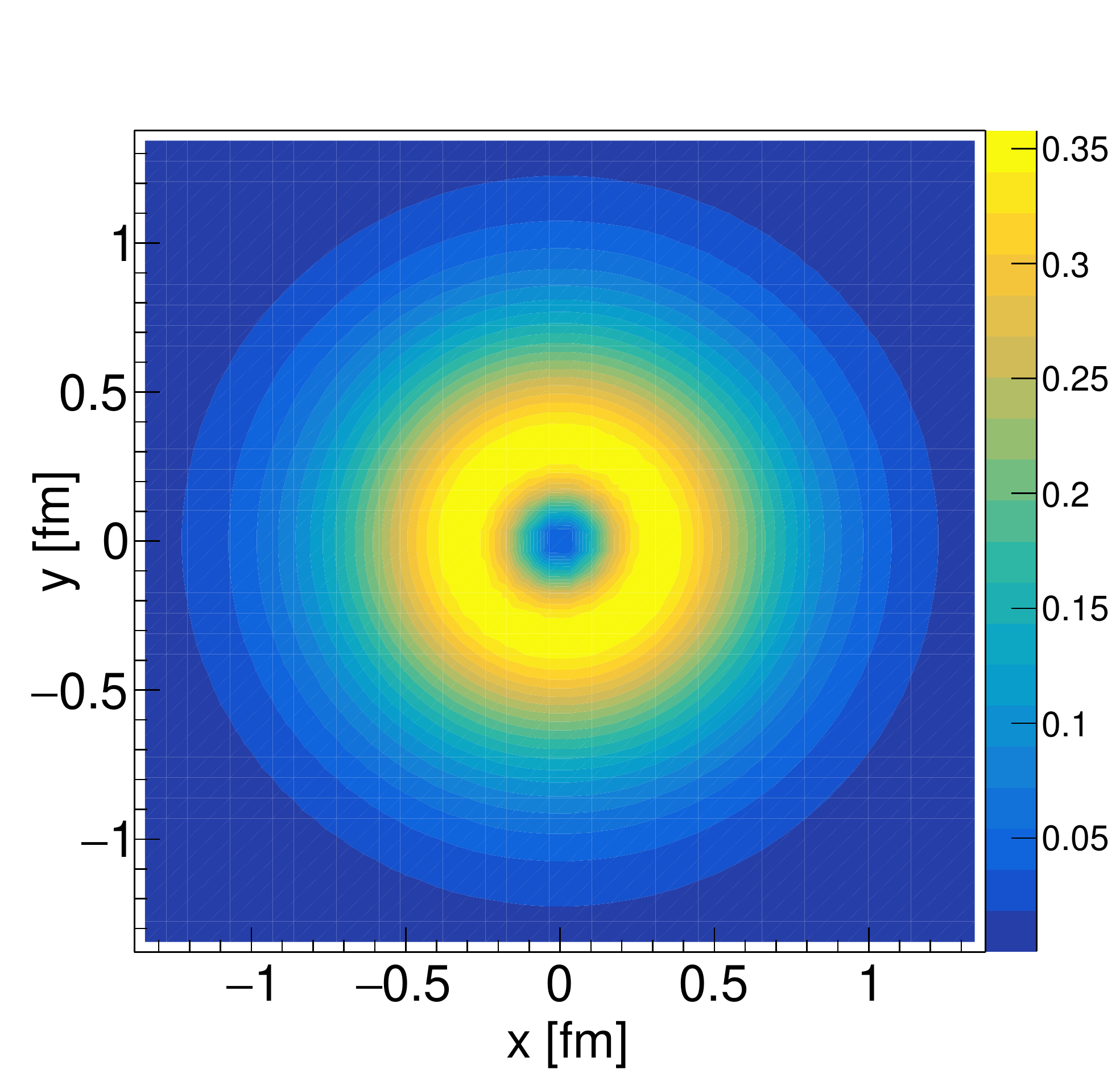}}
	\caption{(Color online) Normalized radial 1-D (a) and 2-D (b) mass distributions of proton $r^2\rho(r)/\rho(0)$ given by the results based on mass radius from experimental data \cite{GlueX:2019mkq,Kharzeev:2021qkd,Wang:2021dis}. The highest values regions for (a) and brightest yellow regions for (b) indicate the maximum mass with large numbers of valence quarks and gluons. Our results are almost consistent with the results of the chiral quark soliton model calculations given in Ref. \cite{Goeke:2007fq}}
	\label{fig:distribution}
\end{figure}

Thus far we have discussed the various radii of proton that are currently available to inform the study of proton structure, and the multi-shell structures we proposed based on these radii. We will now briefly discuss the results we got. The first point is that the mass radius obtained from previous work \cite{Kharzeev:2021qkd,Wang:2021dis} gives us information about the proton mass distribution. It is very important to our understanding of the proton structure, where the mass radius is almost the central value of the axial and gluon radius. This result might be explained by the dominant role of mass inside proton which are valence quarks and gluons contribution (See FIG. \ref{fig:distribution}). 
The mechanical properties inside proton are determined by quarks and gluons. From our analysis in Sec. \ref{sec:multilyer}, we show the multi-shell image of proton by various radii. This internal structure can explain the proton pressure and shear force distribution. Most of the mass of proton is concentrated in the mass radius region, which is the valence quarks and gluons region shown in the bright yellow area in FIG. \ref{fig:2d}. These valence quarks and gluons provide repulsive pressure inside proton, along the radial direction up to the mass radius. Beyond the mass radius there are dispersed pion clouds consisting of gluon-generated sea quarks that provide the confining pressure. It's the combination of confining and repulsive pressure that keeps the proton in a stable state. The physical images we provide are consistent with the results of the chiral quark soliton model to a certain extent \cite{Goeke:2007fq,Polyakov:2018zvc}.

\section{Experimental evidences and Phenomenological explanations}
\label{sec:exp}
\begin{figure}[H]
	\centering
	\includegraphics[width=0.6\textwidth]{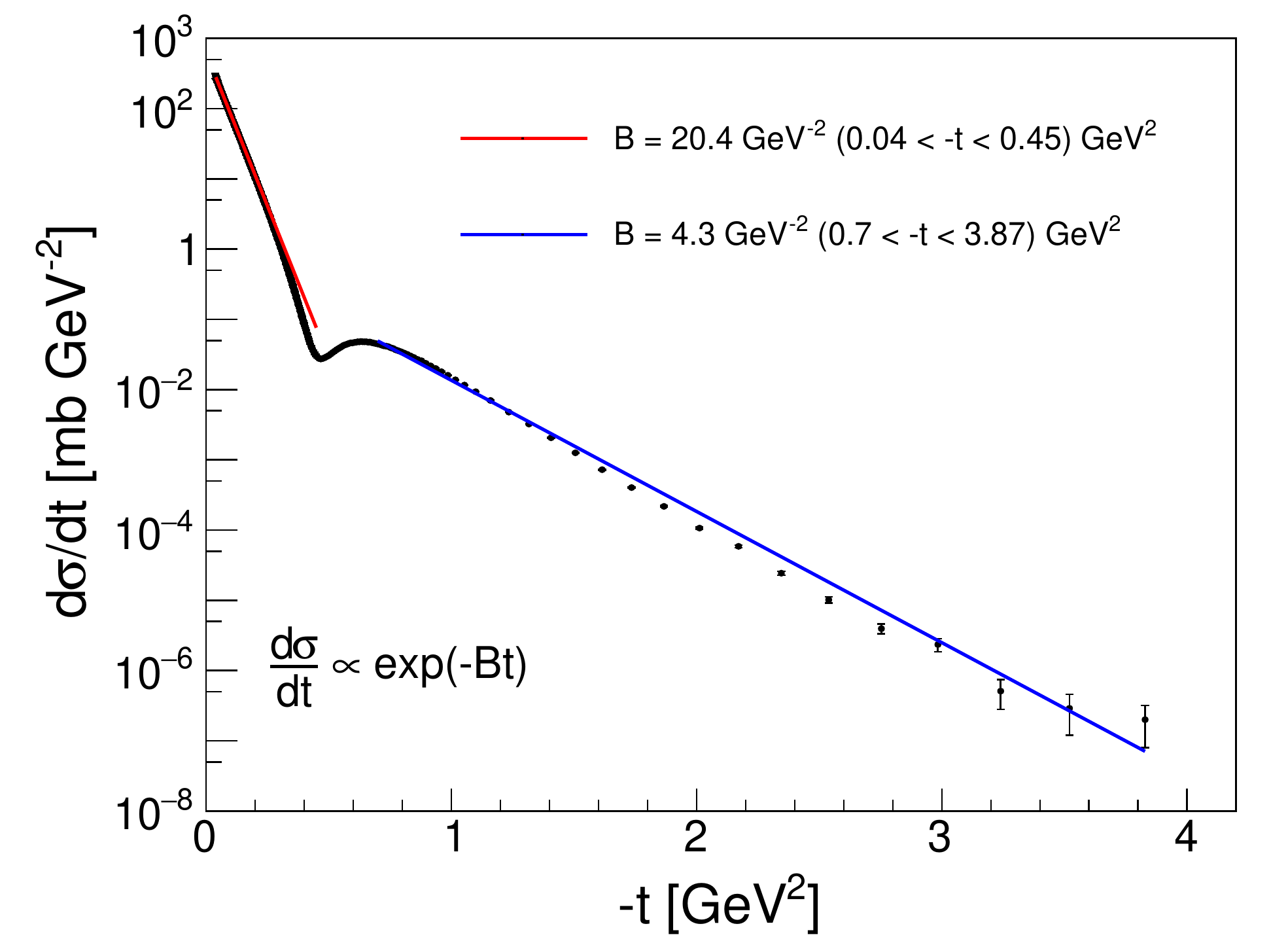}
	\caption{(Color online) Differential elastic cross section $\mathrm{d}\sigma/\mathrm{d}t$ at $\sqrt{s}=13$ TeV \cite{TOTEM:2018hki}. The four-momentum transfer square corresponding to dip from differential cross section distribution is at $-t=0.47$ GeV$^2$. Different slopes are obtained by using the exponential form to fit in different $-t$ ranges. The red and blue solid line represent different slopes \cite{Dremin:2019swd}.
	}
	\label{fig:totem2019}
\end{figure}

Many interesting approaches, both experimental and theoretical, have been given by different researchers in physics to the exploration of the internal structure of the proton. High-energy proton-proton collision experiments always give us novel results. In 2019, the information of high-energy proton-proton elastic scattering experiment by the TOTEM Collaboration can be seen: a dip-like structure exists in the differential cross section with the distribution of four-momentum
transfer squared $-t$ \cite{TOTEM:2018hki}. The results of differential cross section are shown in FIG. \ref{fig:totem2019}. 
One of the obvious points in FIG. \ref{fig:totem2019} is the different slopes on the left and right sides of the dip. Many interpret this theoretically as the imaginary part of the scattering amplitude being zero. One tends to think that the different slopes on difference sides of this ``dip" is understood to a new level inside proton (See Ref. \cite{Islam:1974ac,Islam:2005by,Islam:2008sfk,Dremin:2018urc,Dremin:2019pza,Dremin:2019swd} and their included references), just as Rutherford discovered the existence of a core inside atom over a hundred years ago. The research of odderon contribution to dip can be found in the literature \cite{Csorgo:2018uyp}. The results of the mass radius extraction indicate that the mass radius of proton is significantly smaller than the charge radius \cite{Antognini:2013txn,ParticleDataGroup:2020ssz}. Combined with the high-energy proton-proton collision experiments, we think there have the multi-shell structure inside proton.

As we mentioned in the previous section, the distributions of the various components within the proton is energy-scale-dependent. Our previous works have also demonstrated that the internal mass composition of proton varies at different energy scales \cite{Wang:2019mza,Kou:2021bez}. The proton velocity approaches the speed of light at the high energy experiment, and the quarks inside proton split into quark-antiquark pairs under this condition. Based on the researches about the proton mass decomposition, proton mass could be divided in four parts, quark energy $M_q$, gluon energy $M_g$, quark mass $M_m$ and trace anomaly (quantum anomalies energy) $M_a$. Quantum anomalies energy, $M_a$, is reflected in the gluon dynamics behavior inside proton as part of the source of proton mass \cite{Ji:1995sv,Ji:2021pys}. We use the results in \cite{Kou:2021bez} and give specific images to describe the structure inside proton. We consider the trace anomaly of gluons to be energy-scale-dependent, so the evolution of trace anomaly with energy scale can be written as
\begin{equation}
	\frac{\mathrm{d} T_{a}}{\mathrm{d} \mu^{2}}=\frac{A}{\mu^{4}} T_{a},
	\label{eq:evolution}
\end{equation}
where $T_{a}=M_{a} / M_{N}=0.25(1-b)$ is the fraction of the QCD trace anomaly to the proton mass and the separation of mass defined in \cite{Ji:1995sv}. According to the evolution equation of the trace anomaly with energy scale we can obtain the decomposition of proton mass at different scales.

\begin{figure}[H]
	\centering
	\includegraphics[width=0.7\textwidth]{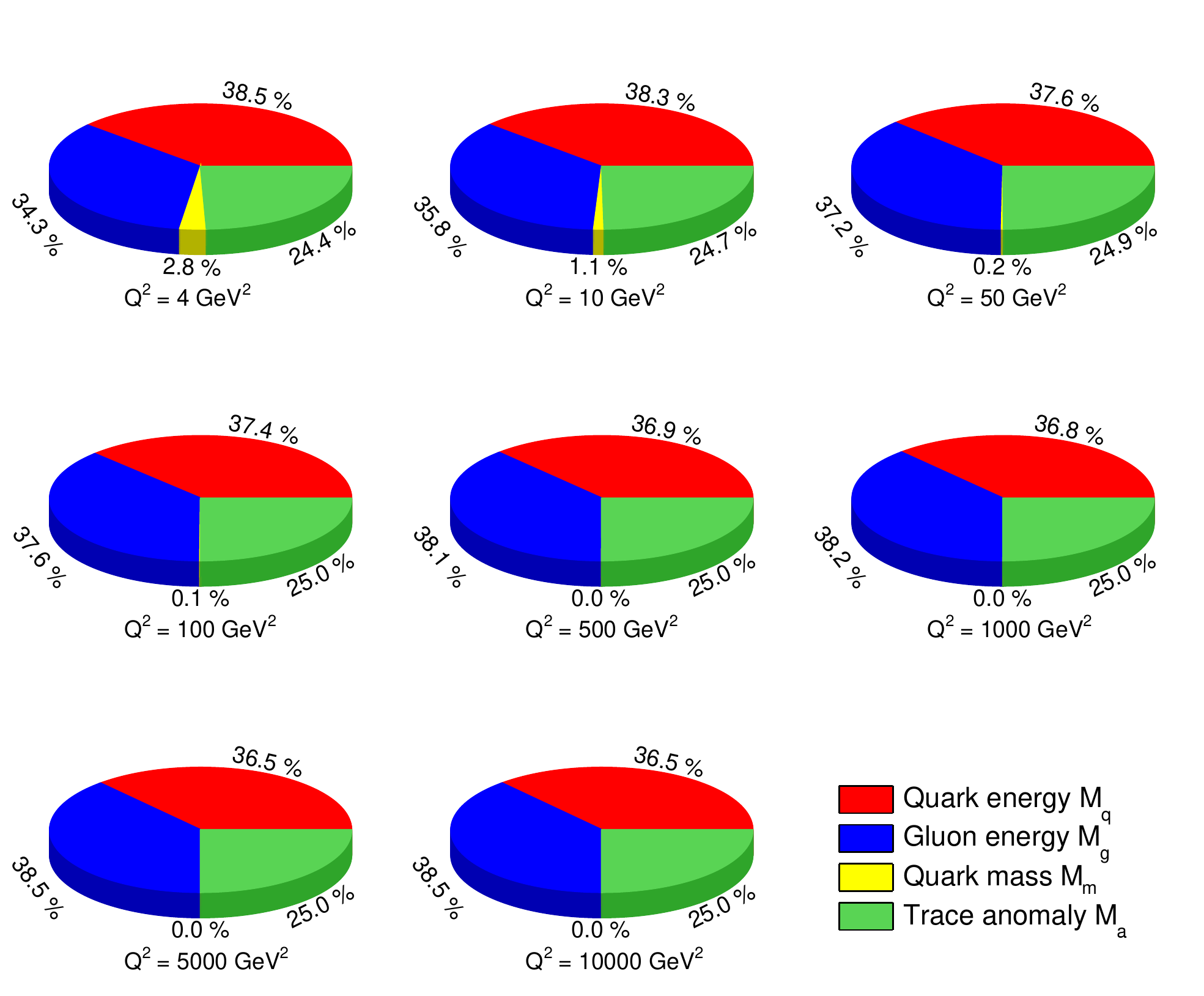}
	\caption{(Color online) Mass decomposition at high energy scale $Q^2 =$ 4, 10, 50, 100, 500, 1000, 5000 and $10000$ GeV$^2$. The results are based on \cite{Ji:1995sv,Wang:2019mza,Kou:2021bez}.
	}
	\label{fig:massdecomposition}
\end{figure}

\begin{figure}[H]
	\centering
	\includegraphics[width=0.7\textwidth]{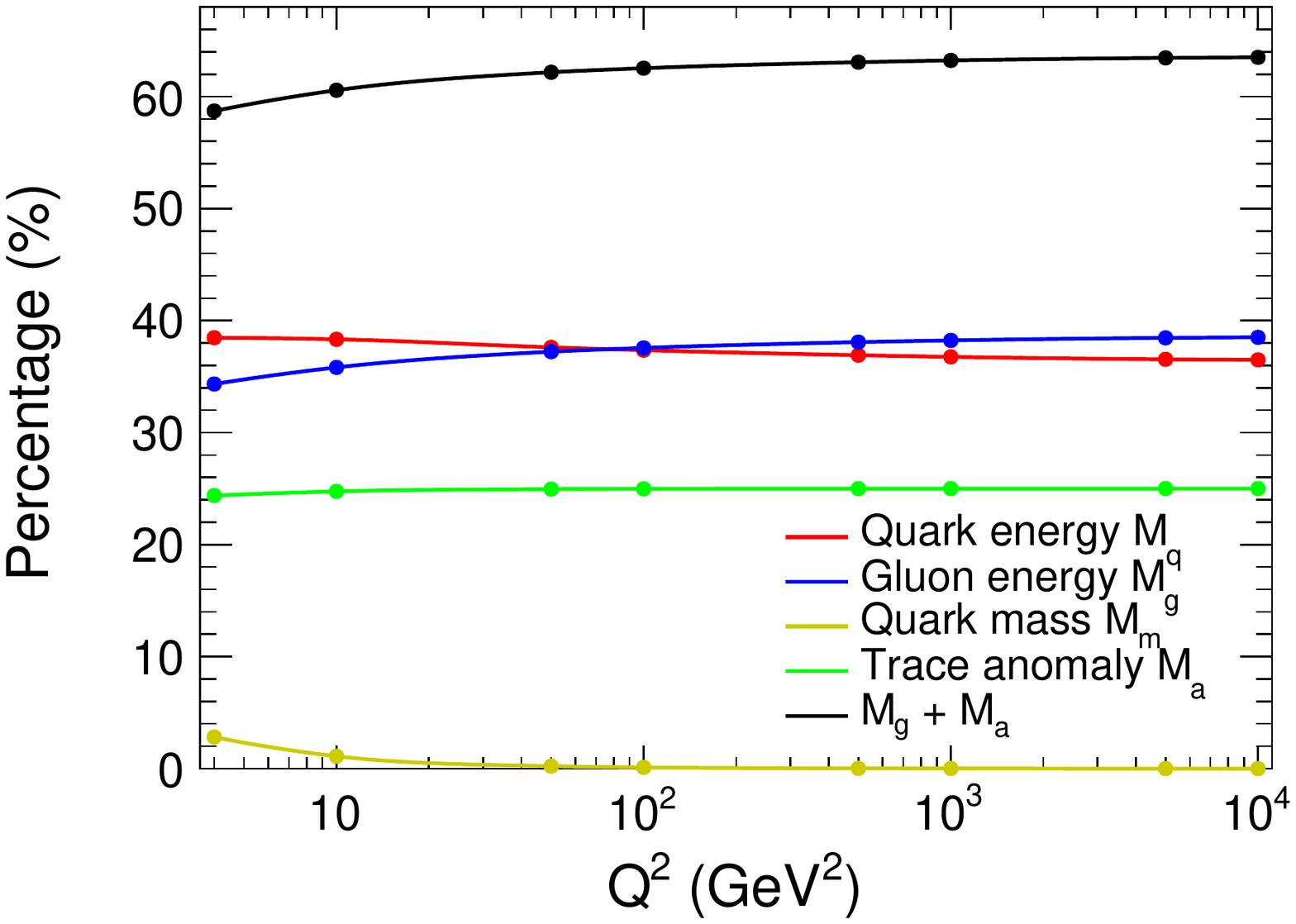}
	\caption{(Color online) Four parts of mass decomposition corresponding FIG. \ref{fig:massdecomposition}. The black points and solid line represents the total contribution of gluon inside proton. Gluons account for roughly two-thirds of the proton in the high-$Q^2$ case.
	}
	\label{fig:percentage}
\end{figure}

Figures \ref{fig:massdecomposition} and \ref{fig:percentage} reveal that at high energy the valence quarks within the proton almost disappear and transfer into gluons. In this case there is a large number density of gluons and many generated sea quark pairs (meson clouds) inside proton. We now focus on the proton-proton collision experiment of the TOTEM Collaboration. At the center-of-mass energy of $\sqrt{s} = 13$ TeV, the fast proton has led to the production of a large number of gluons, and gluons are recombined in a small volume due to the Lorentz contraction effect at high velocity condition. As a result, the gluon density is very high. At this moment the scattering between two proton beams occurs on the gluon ``shell" which can be treated as rigid spheres. The high-$Q^2$ $pp$ collisions occurred in a flash and proton can be divided into shells this moment--hard gluon and soft meson clouds. Since the outer ``shell” of the proton is a ``soft" meson clouds made by sea quark pairs, the high-energy proton can pass right through. However, due to the appearance of the high-energy scalar gluons we mentioned earlier, the inner core of the proton is a rigid ball made by compact gluons. Therefore, a large angle elastic scattering occurs, and an inflection point appears in FIG. \ref{fig:totem2019}. It is similar to the Rutherford scattering experiment that discovered the nucleus \cite{Rutherford:1911zz}.

\section{Conclusions and Discussions}
\label{sec:discussion}

In this work, we propose a new shell structure of the proton. Based on the mass distribution of proton and the mass radius, we give information about the mass decomposition structure of proton at different energy scales. We divided the proton into three shells (low-$Q^2$) and two shells (high-$Q^2$) according to the radii corresponding to the individual interactions within proton. Among them, the charge radius as well as the mass radius we consider to be independent of the energy scale. In previous sections we also gave the reasons for our proposed proton shell structure as well as the phenomenological explanations. Therefore the shell structure of the proton should also be closely related to the results of the decomposition of proton mass. 

The proton structure studies belong to the QCD non-perturbative energy region physics. Both high-energy DIS as well as low-energy vector meson photo-production experiments can be used to study the evolution inside proton. Due to the normalization condition, the momentum fractions carried by the parton of the proton have to be equal to 1 after integration by Bjorken scale $x_B$, which means that the gluon and quark parts can transform into each other. But the total mass of the proton or the reflection of the gravitational effect is constant, just the mass distribution will be different due to the evolution of the parton. There are already relatively accurate experimental results for photon-proton scattering at high or low energy. The Electron-Ion colliders in the United States \cite{Accardi:2012qut,AbdulKhalek:2021gbh} and China \cite{Chen:2018wyz,Chen:2020ijn,Anderle:2021wcy} will definitely offer more precise experimental data for the internal structure of proton and hence test this proton shell model in future.

\begin{acknowledgments}
This work is supported by the Strategic Priority Research Program of Chinese Academy of Sciences
under the Grant NO. XDB34030301.
\end{acknowledgments}

\bibliographystyle{apsrev4-1}
\bibliography{refs}

\end{document}